%% file: ms.tex
\documentclass[10pt,conference]{IEEEtran}
\IEEEoverridecommandlockouts
\def\BibTeX{{\rm B\kern-.05em{\sc i\kern-.025em b}\kern-.08em T\kern-.1667em\lower.7ex\hbox{E}\kern-.125emX}}

\usepackage{settings}
\usepackage{listings}

\newboolean{9pager} 
\setboolean{9pager}{true}
\newif\ifshort
\ifthenelse{\boolean{9pager}}
{
\shorttrue 
}
{}

\begin{document}

\title{
Physical Wireless Resource Virtualization for Software-Defined Whole-Stack Slicing  
\ifthenelse{\boolean{9pager}}
{}
{
}
\thanks{This work is supported in part by the NSF awards 1827211 and 1821962.}
}

\author{
\IEEEauthorblockN{Matthias Sander-Frigau, Tianyi Zhang, Hongwei Zhang, Ahmed E. Kamal, Arun K. Somani}
\IEEEauthorblockA{\textit{Department of Electrical and Computer Engineering, Iowa State University} \\
\{msfrigau,tianyiz,hongwei,kamal,arun\}@iastate.edu
}
}

\maketitle

\input{abstract.tex}
\input{introduction.tex}

\input{relatedWork}
\input{preliminaries}

\input{pvran}

\input{measurement.tex}

\input{conclusion.tex}

\ifthenelse{\boolean{9pager}}
{}
{ \input{ack}  }

\bibliographystyle{IEEEtran}
{
\bibliography{references-DNC,references}
}

\end{document}

%% file: abstract.tex
\begin{abstract}
Radio access network (RAN) virtualization is gaining more and more ground and expected to re-architect the next-generation cellular networks. 
    Existing RAN virtualization studies and solutions have mostly focused on sharing communication capacity and tend to require the use of the same PHY and MAC layers across network slices. This approach has not considered the scenarios where different slices require different PHY and MAC layers, for instance, for radically different services 
    and for whole-stack research in wireless living labs where novel PHY and MAC layers need to be deployed concurrently with existing ones on the same physical infrastructure. 
To enable whole-stack slicing where different PHY and MAC layers may be deployed in different slices, we develop \emph{PV-RAN}, the first open-source virtual RAN platform that enables the sharing of the same SDR physical resources across multiple slices. 
    Through API Remoting, PV-RAN enables running paravirtualized instances of OpenAirInterface (OAI) at different slices without requiring modifying OAI source code. 
PV-RAN effectively leverages the inter-domain communication mechanisms of Xen to transport time-sensitive I/Q samples via shared memory, making the virtualization overhead in communication almost negligible. 
    We conduct detailed performance benchmarking of PV-RAN and demonstrate its low overhead and high efficiency. We also integrate PV-RAN with the CyNet wireless living lab for smart agriculture and transportation.

\end{abstract}

%% file: introduction.tex
\section{Introduction}

Recent advances in 5G virtualization \cite{alvarez2019edge}, 
through the use of Network Function Virtualization (NFV) and Software Defined Networking (SDN), have led to the development of the concept of Network Slicing  \cite{zhang2017network}. 
Although there is no consensus on the strict definition of a Network Slice, it generally refers to an end-to-end logical portion of the network resources. IETF defines Network Slicing as the collection of a set of technologies to create specialized and dedicated logical networks as a service (NaaS) in support of network service differentiation and meeting the diversified requirements from vertical industries \cite{peng2019packet}. Thus, network slicing has the capability for the network to provide a multitude of services that demand very specific requirements in terms of latency, reliability, and bandwidth.
According to telecommunications Standard Developing Organizations (SDOs), network slicing is a key enabler for accommodating diverse 5G services such as ultra-reliable and low-latency communications (URLLC), extreme mobile broadband (xMBB), and massive machine-type communication (mMTC).

Critical to network virtualization and slicing is the virtualization and slicing of radio access networks (RANs) in addition to core networks 
and edge/cloud resources. Despite much progress in RAN virtualization and slicing, existing work has mostly focused on sharing communication capacity, and the different slices tend to use the same PHY and MAC layers \cite{FlexRAN,esmaeily2020cloud,wirelessVirtualization:survey}. 
    Nonetheless, the heterogeneous requirements of 5G and beyond wireless services may well demand different PHY and MAC layers. For instance, optimal URLLC and mMTC solutions may well differ in their PHY and MAC layers due to the radically different requirements that URLLC and mMTC pose to communication timeliness, reliability, throughput, and energy efficiency. 
Therefore, there exists the unmet need for RAN virtualization solutions that enable different slices to run different PHY and MAC layers. For convenience, we call this type of RAN virtualization \emph{whole-stack slicing}. Whole-stack slicing is not only important for production 5G-and-beyond systems, it is also important for enabling whole-stack research and innovation in wireless living labs where novel solutions in the PHY and MAC layers can be experimented together with existing solutions \cite{CyNet,UCS-ICII18}. 

\begin{figure}[!htbp]
    \centering
    \includegraphics[width=\linewidth]{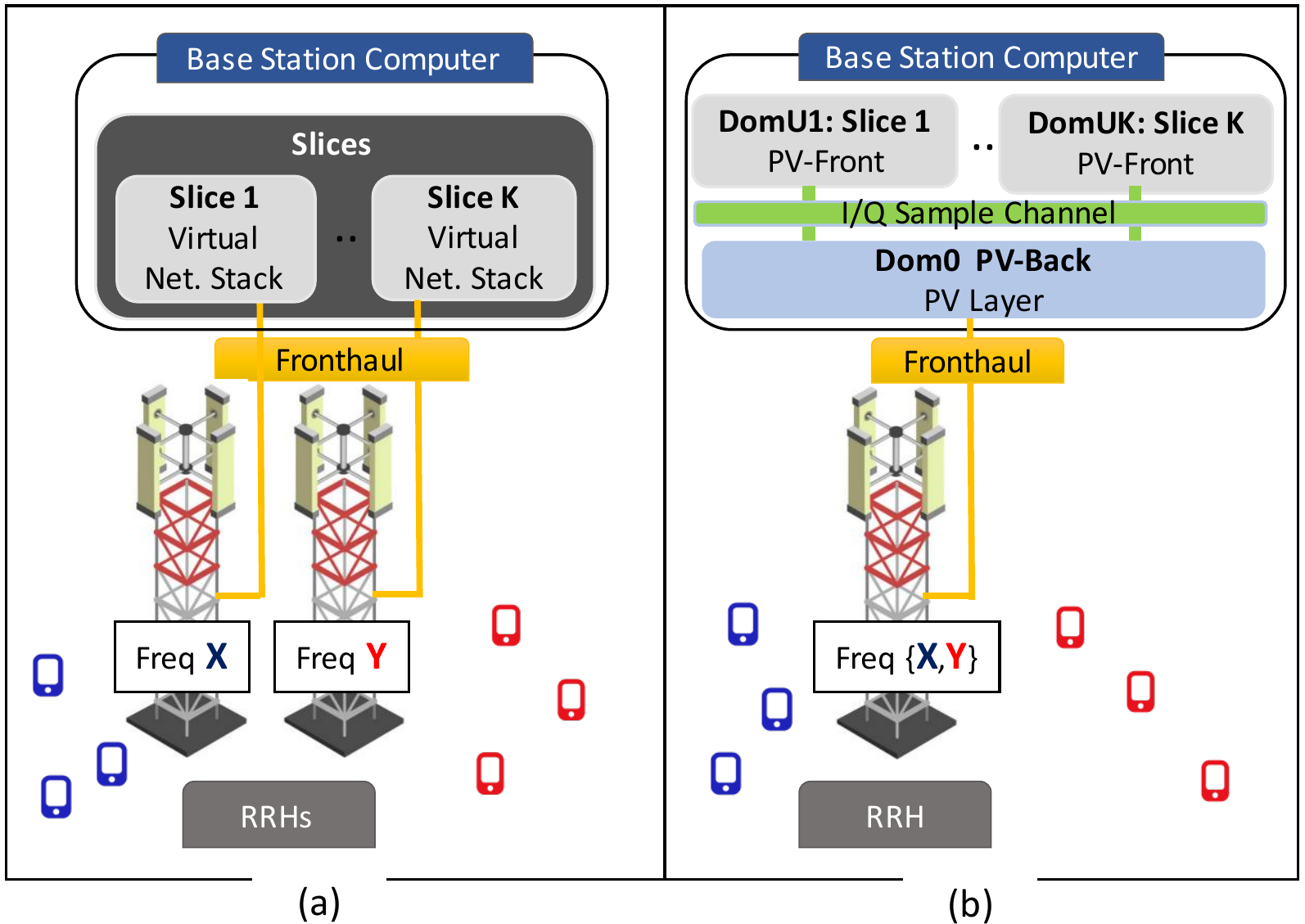}
    \caption{(a) Traditional C-RAN where the fronthaul connects slices (with virtualized network stacks) to their dedicated remote-radio-heads (RRHs). 
    (b) PV-RAN where the fronthaul connects a shared RRH to multiple slices.}
    \label{fig:archcynet}
    \vspace*{-0.1in}
\end{figure}

To enable the use of different PHY and MAC layers at different RAN slices, we need to enable the sharing of the physical wireless radios, a.k.a$.$ remote radio heads (RRHs), across slices. In this study, we use USRP software-defined-radios (SDRs) \cite{USRP} as the physical radios, and we propose the \emph{PV-RAN} platform that virtualizes the RAN SDRs. PV-RAN uses the Xen hypervisor \cite{Xen} as the base virtualization platform for the virtual RANs, and its current implementation assumes that the open-source wireless software platform OpenAirInterface (OAI) \cite{OAI} is used as the protocol implementation platform. As shown in Figure~\ref{fig:archcynet}, 
PV-RAN spans two main components of each RAN: 
\begin{itemize}
    \item \emph{PV-Back} in the Dom0 of the base station computer: it interacts with the SDR and implements the logic necessary to virtualize and multiplex SDR radio resources. 
    \item \emph{PV-Front} in each DomU of the base station computer: it runs the full network protocol stack (e.g., from PHY to PDCP in LTE) with a modified version of the USRP low-level interface of OAI that accounts for SDR radio resource virtualization.
\end{itemize}
Compared to the C-RAN architecture that makes use of at least one radio per slice, the PV-RAN platform can support multiple slices using the same radio. 
    
    The contributions of this work are as follows: 
\begin{itemize}
    \item We present the first open-source design and implementation of the virtualization of physical SDR resources to enable whole-stack slicing. Besides supporting whole-stack slicing in production systems, PV-RAN allows researchers to prototype new cellular network PHY and MAC layers using exclusively open-source software: PV-Back and PV-Front use Xen paravirtualization and its inter-domain communication mechanism to transport I/Q samples, and the virtualized slices use OAI as the wireless network protocol stack. 
    
    \item PV-RAN enables running paravirtualized instances of OAI without requiring modifying OAI source code. This is accomplished through a novel \textit{API Remoting} method that intercepts UHD library function calls from paravirtualized OAIs in DomUs and redirect them to the privileged domain Dom0 where the library call is executed and forwarded to the SDRs. 
    
    \item PV-RAN ensures slice isolation in shared SDR access, and it features novel and effective techniques such as lightweight, in-band timestamping synchronization between PV-Back and PV-Front. 
    
    \item We perform detailed performance benchmarking of PV-RAN, and demonstrate that  the inter-domain communication channel of Xen is a fast and efficient interface for the transport of time-sensitive I/Q samples. In comparison to OAI 
    running in bare-metal computers without virtualization, the overhead due to virtualization is insignificant. 
    
    \item We have integrated PV-RAN with the software-defined cyberinfrastructure CyNet~\cite{CyNet}, the first field-deployment of open-source, virtualized cellular platforms for smart agriculture and transportation. 
\end{itemize}

The rest of this paper is organized as follows. We discuss related work in Section~\ref{sec:relatedWork}. in Section~\ref{sec:preliminaries}, we present the system model and introduce the Xen hypervisor and CyNet wireless living lab. 
    We present the design and implementation of PV-RAN in Section~\ref{sec:designImpl}. 
We evaluate the performance of PV-RAN in Section~\ref{sec:evaluation}. We share concluding remarks in Section~\ref{sec:conclusion}. 

%% file: relatedWork.tex
\section{Related Work} 
\label{sec:relatedWork}

\ifthenelse{\boolean{9pager}}
{
Virtualized RAN (vRAN) centralizes and virtualizes a baseband-unit pool (vBBU) at a strategic  location (e.g., central office), where the vBBU can be deployed on 
commercial-off-the-shelf (COTS) servers rather than proprietary hardware. vRAN  allows operators to rapidly expand the capacity and coverage of the network, thus minimizing capital expenditure (CAPEX) and operating expenditure (OPEX)~\cite{suryaprakash2015heterogeneous}. 
}{
\subHeadingS{RAN virtualization.} 
The idea of RAN virtualization started with the Cloud-RAN (C-RAN) initiative, which was initially introduced in 2009 by the China Mobile Research Institute~\cite{wang2010c}. The basic idea of C-RAN was to decouple the radio elements (RRU/RRH) from the baseband processing units (BBU), where BBUs are connected to RRHs via CPRI fiber-based fronthaul links. In C-RAN, instead of having BBUs at the cell site, BBUs are shifted towards a central datacenter. Conceptually, the centralization of BBUs allowed a more flexible resource coordination between multiple cell sites, making it possible to implement joint processing and scheduling solutions~\cite{kassab2018coexistence} that mitigate inter-cell interference, and to perform intelligent scaling of resources that reduces deployment cost when capacity demand increases. C-RAN has played a substantial role in reshaping the conventional RAN architecture to bring higher scalability and flexibility of further deployment of RRHs~\cite{wang2014potentials}.

Due to this centralization of baseband processing, it became natural to adopt principles from SDN/NFV to virtualize baseband processing resources in order to dynamically allocate resources on-demand depending on time-varying traffic conditions. This evolution towards virtualization is know as the virtualized-RAN (vRAN). A vRAN centralizes and 
virtualizes a BBU pool (vBBU) at a strategic location (i.e., central office), where vBBU can be deployed on 
commercial-off-the-shelf (COTS) servers rather than proprietary hardware which allows operators to rapidly expand the capacity and coverage of the network, thus minimizing capital expenditure (CAPEX) and operating expenditure (OPEX)~\cite{suryaprakash2015heterogeneous}. 
}
Garcia et al.~\cite{garcia2018fluidran} proposed an analytical framework for vRAN called FluidRAN that optimizes the placement of vRAN functions jointly with the routing policy according to network and computing resources. 
Nikaein et al.~\cite{nikaein2017towards} have considered potential vRAN architectures that could meet real-time deadlines and intensive computational and I/O requirements. 

With the availability of SDR platforms and OAI and the growing interest toward virtualization, researchers can prototype low-cost cellular 
networks and evaluate the performance of different virtualized RAN architectures~\cite{kaltenberger2020openairinterface}.
    In~\cite{tran2017understanding}, Tran et al. evaluated a C-RAN testbed based on the OAI RRH/BBU split and virtualized the OAI BBU-pool in VMWare virtual machines. Their study provides interesting insight into the computational requirements of vBBU under different radio resource configurations. The H2020 5G-PPP SliceNet project~\cite{sanchez20205gtoponet} introduced a slice-friendly virtualized 5G RAN where the Central Unit (CU) and Distributed Unit (DU) run in LXC containers. 
In~\cite{trindade2019c}, 
Trindade et al$.$ containerized OAI in Docker and used Kubernetes to orchestrate the RRH and BBU containers. Nikaein et al$.$~\cite{nikaein2015processing} investigated the performance of virtualized BBUs under containers (Docker, LXC) and KVM virtual machines. 


Esmaeily et al.~\cite{esmaeily2020cloud} recently introduced a 
testbed 5GIIK capable of performing E2E network slicing. They rely on the ETSI NFV MANO framework to manage and orchestrate the resources required for creating, managing, and delivering services through different slices. 
    The FlexRAN~\cite{FlexRAN} platform allows the programmable control of the underlying RAN infrastructure via virtual control functions, and it adds a virtualization layer over the RAN infrastructure to enable communication capacity allocation across slices. 

The aforementioned studies have touched on various aspects of RAN virtualization, but they have not considered virtualizing the physical wireless resources for whole-stack slicing. 
    The work closest to PV-RAN is that of the Wireless Spectrum Hypervisor~\cite{de2020baseband}. It considered multiplexing several OFDM signal streams through a shared SDR. It cannot support slices running non-OFDM PHY layers; the spectrum hypervisor requires moving the OFDM modulation component from the user slices to the hypervisor layer, thus not being transparent to the slice users; it also uses ZeroMQ (a.k.a$.$ ZMQ) bus for communication between user slices and the hypervisor, which introduces much higher overhead than the lightweight inter-domain communication mechanism adopted by PV-RAN, as we show in Section~\ref{sec:evaluation}. 

\ifthenelse{\boolean{9pager}}
{}
{
\subHeading{Shared-memory inter-domain communications.} 
By bypassing the traditional TCP/IP protocol stack, shared memory I/O channels have been explored to enable efficient inter-domain communications (IDC) in Xen-based virtualized systems. 
XenLoop~\cite{wang2008xenloop} is a shared memory IDC mechanism that minimizes latency and maximizes bandwidth of inter-domain networking applications. It allows existing networking applications to bypass traffic redirection to Dom0. XenLoop intercepts traffic at layer 3 to deliver high-performance communication across domains. It preserves user-level transparency and supports  guest domains migration. 
    MMNet (Fido)~\cite{burtsev2009fido} is another layer 3 IDC mechanism that eliminates data copies and hypervisor calls (hypercalls) in the data path to provide fast IDC performance aimed at large-scale distributed applications. Other prior works have addressed the inter-domain communication performance degradation by leveraging Xen shared memory mechanism such as  XenSocket~\cite{zhang2007xensocket}, XWAY~\cite{kim2008inter}, IVC~\cite{huang2007virtual},  XenVMC~\cite{ren2012fast} and IVCOM~\cite{bai2013high}.
For a comprehensive list of features and design choices of Xen IDC mechanisms, we redirect the readers to the survey by Ren et al.~\cite{ren2016shared}. 
    These mechanisms are intended for TCP/IP based inter-domain networking applications only, and they do not address the unique needs of inter-domain communication for I/Q sample streaming and physical wireless resource virtualization. 

} 

%% file: preliminaries.tex
\section{Preliminaries} \label{sec:preliminaries}

\begin{figure}[!htbp]
    \centering
    \includegraphics[width=\linewidth]{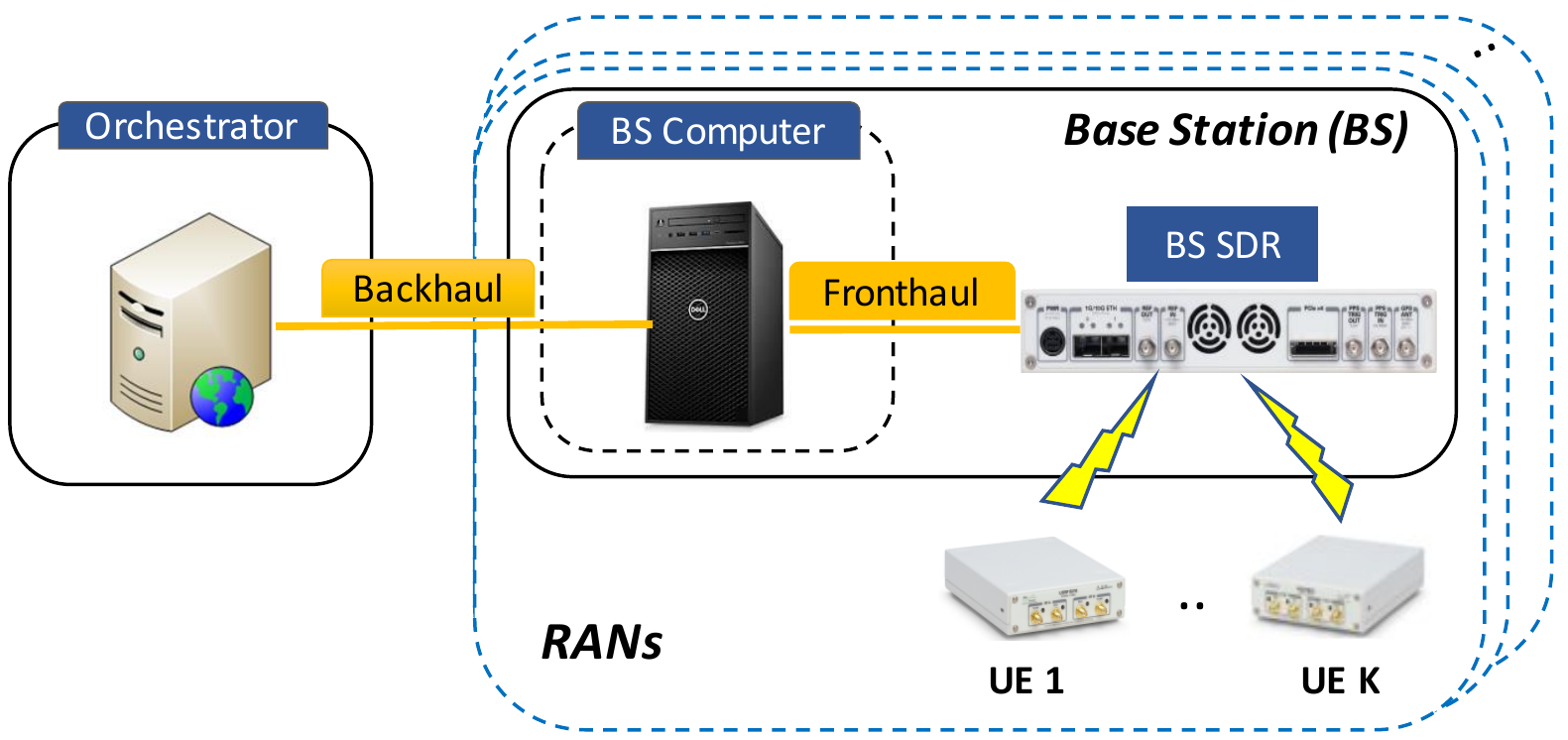}
    \caption{System model}
    \label{fig:systemModel}
    \vspace*{-0.1in}
\end{figure}

\subHeadingS{System model.}
As shown in Figure~\ref{fig:systemModel}, 
the system consists of an orchestrator and a set of RANs. Each RAN has a base station (BS) and a set of User Equipment (UEs), with each BS consisting of a SDR and a computer. The SDR serves as the remote radio head (RRH), and the computer runs the cellular network stacks realized as software. 
    A network slice consists of a RAN slice at one or more RANs, and each RAN slice runs a cellular network stack implemented using OpenAirInterface (OAI) \cite{OAI} with potentially different PHY and MAC layers. 
We consider Frequency-Division-Multiplexing (FDM) in RAN slicing such that different slices of a RAN operate in non-overlapping frequency bands. The specific frequency division strategies at different RANs could be different or the same, and this is coordinated through the central orchestrator and defined by software. 
    Within this context, we study the systems issues of how to virtualize the BSes so that a BS SDR can be shared among multiple slices of the associated RAN to support whole-stack slicing (i.e., with each slice running potentially different PHY and MAC layers). 

For simplicity of exposition, here we assume that each RAN only has one BS SDR, but the virtualization solution PV-RAN is readily extensible to the case where each RAN may have multiple SDRs. 
    As we will explain shortly, a key task of PV-RAN is multiplexing/demultiplexing I/Q samples between the BS SDR and the network stacks of different RAN slices. Therefore, the architecture considered here resembles the RAN functional split Option 8 as defined by 3GPP \cite{3gpp-functionalSplit}. The PV-RAN design, however, is readily extensible to other functional split options such as Options 7-1 and 7-2 where frequency-domain I/Q samples need to be multiplex-/demultiplexed between the BS SDR and different network stacks running on the BS computer. 
Similarly, the PV-RAN framework can be extended to network slicing strategies involving Time Division Multiplexing (TDM), and the network stacks could be based on other open-source cellular software platforms such as srsLTE. These are all interesting topics for future exploration, but their detailed studies are beyond the scope of this work. 

\subHeading{Xen hypervisor.}
Considering the security benefits of virtual machines (VMs) as compared with containers \cite{LightVM}, as well as the facts that VMs can be made lightweight \cite{LightVM} and that paravirtualization enables efficient I/O operations, we use the Xen hypervisor \cite{Xen} as the base virtualization platform in our PV-RAN implementation.\footnote{Other paravirtualization platforms such as KVM may also be used, but detailed study is beyond the scope of this work. In addition, thanks to its microkernel architecture, Xen provides a greater degree of isolation between the hypervisor and domains than KVM does.} 
In particular, we use Xen to manage the BS computer VMs used by individual slices of a RAN. 
    Xen hypervisor runs directly on top of the hardware and provides abstraction and isolation required for the operation of multiple guest OSes hosted on the same physical computer. A Xen-based virtualization system consists of the hypervisor and the privileged domain Dom0 that has direct hardware access and can manage one or more unprivileged guest domains (DomUs).

In Xen, Dom0 has privileged access to hardware (e.g., SDR) while DomUs are typically not allowed to access hardware directly. However, Xen hypervisor supports an 
I/O virtualization mechanism, known as the \textit{split-driver model}, allowing DomUs to access hardware through the use of virtual device drivers.
    In particular, a front-end device driver lies in a DomU and a back-end device driver resides in Dom0 and talks to the native device driver. The front-end and back-end drivers use I/O ring buffers to communicate requests/replies. I/O ring buffers rely on the \textit{grant table} mechanism to share memory pages between domains, and they use \emph{event channels}, an asynchronous notification mechanism, to notify a domain when there are waiting data in the I/O ring buffers. Thus, grant tables and event channels are convenient Xen facilities that allow to create I/O channels to establish bi-directional communication between domains. 
As we will explain in detail in Section~\ref{sec:designImpl}, PV-RAN adopts the split-driver model for efficient I/Q sample streaming between Dom0 and DomUs.

\subHeading{CyNet wireless living lab.} 
CyNet \cite{CyNet} is a field-deployed wireless living lab, consisting of two cellular RANs deployed at the Iowa State University Curtiss Research Farm and Research Park in the City of Ames, Iowa, USA. Each RAN has 2-3 USRP X310s as the BS SDRs and several USRP B210s as UEs. This work is motivated by the need for CyNet to partition its RANs into different slices that employ different PHY and MAC layers to support research and education in wireless, agriculture, and transportation respectively. PV-RAN has been tested and deployed with the CyNet infrastructure.

%% file: pvran.tex
\section{PV-RAN Design and Implementation} \label{sec:designImpl}

In our implementation of PV-RAN, the BS SDR is the USRP X310, and the UE SDR is the USRP B210. For ease of discussion, we will refer to these specific SDRs and related software modules (e.g., the SDR driver UHD) in this section, but the design and implementation strategies of PV-RAN are readily applicable/extensible to other SDR platforms.

\subsection{Architecture} \label{subsec:architecture}

\begin{figure}[!htbp]
    \centering
    \includegraphics[width=\linewidth]{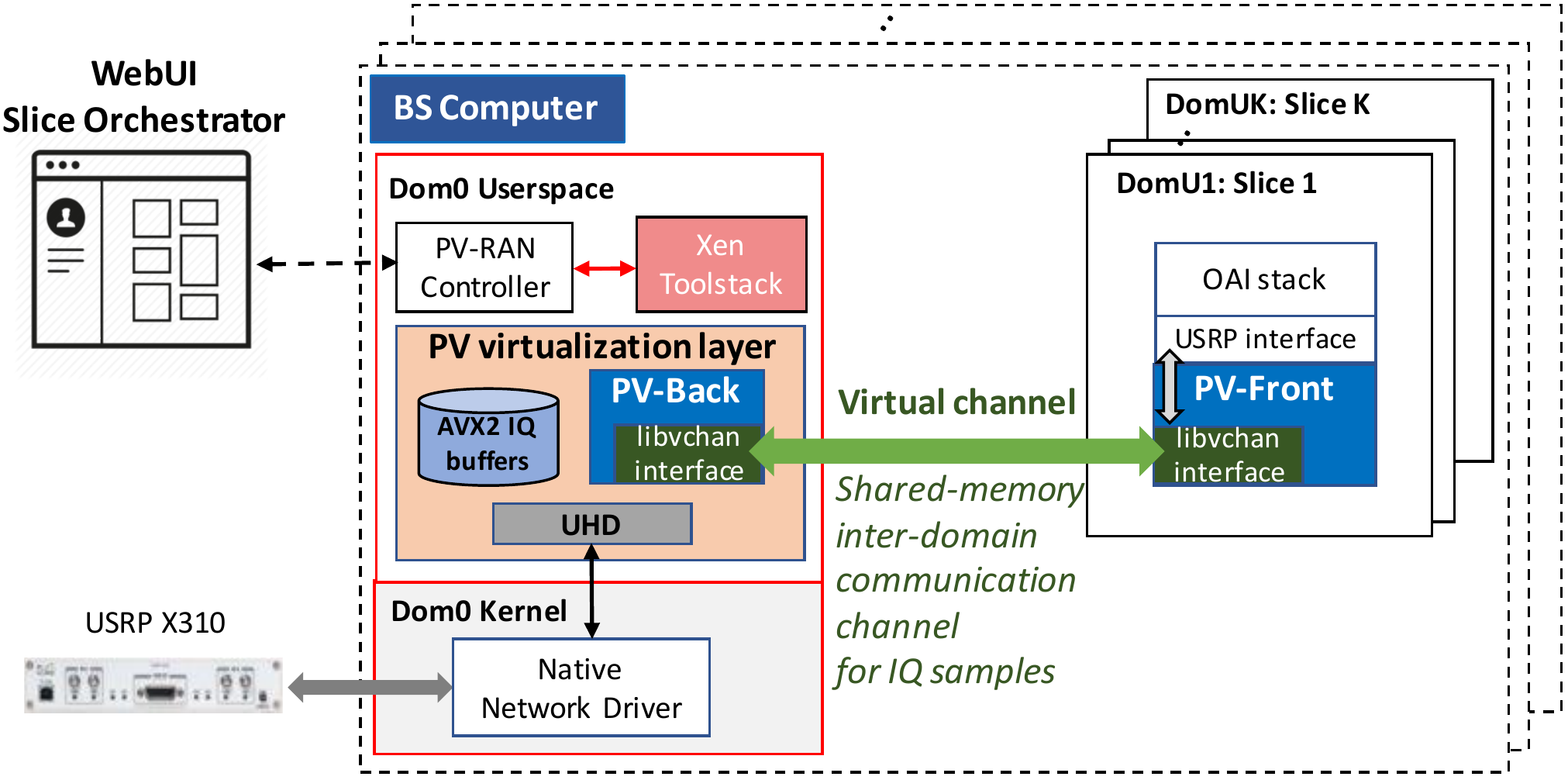}
    \caption{Architecture of PV-RAN}
    \label{fig:archPV-RAN}
\vspace*{-0.1in}
\end{figure}

Figure~\ref{fig:archPV-RAN}
shows the architecture of the PV-RAN. 
    The software of the full OAI network stacks of user slices are deployed in the DomUs of the Xen-virtualized BS computer, and they communicate with the physical SDR via the PV-Front in the individual DomUs; the virtual channel carries the I/Q samples through the Xen libvchan interface; PV-Back is deployed in Dom0 to perform all the baseband I/O operations from/to the USRP X310 via Dom0's kernel native network driver; USRP X310 integrates the antennas and acts as the RRH, and it communicates with Dom0 using the USRP Hardware Driver (UHD). 
The PV-RAN controller in Dom0 acts as a middleware layer between the WebUI Slice Orchestrator and the Xen toolstack \texttt{xl} by enforcing security and control. It enables the WebUI Slice Orchestrator, a Web portal for software-defined orchestration and management of DomUs, to send orchestration messages to the Xen toolstack that controls DomUs. When an end-user requests the creation of a new DomU through the Web portal, an orchestration message is  sent to the PV-RAN controller which will convert it into an \texttt{xl} command, which in turn will create the new DomU user VM. 
\ifthenelse{\boolean{9pager}}
{}
{ In our implementation, the orchestration messages between the PV-RAN controller and the Web portal are exchanged over a ZeroMQ communication socket using the REQ-REP pattern over TCP port 5555. }

In order to run multiple heterogeneous paravirtualized OAI protocol stacks in different DomUs (i.e., user slices), 
each isolated OAI instance must appear to have dedicated access to the SDR. Since only one process can access the USRP Hardware Driver (UHD) at a time, we must decouple RF I/O (a.k.a$.$ baseband I/O) from the PHY layer to allow multiple concurrent OAI instances to simultaneously access the USRP device.  This is achieved by virtualizing the SDR hardware resource 
and delegating all RF I/O operations to a shim layer located in the Xen Dom0 and between the PHY layer of each paravirtualized OAI instance and the actual UHD. This shim layer is called the \emph{PV virtualization layer}. 
The virtualization layer provides the following capabilities: 
\begin{itemize}
  \item \textbf{Slice isolation:} the virtualization layer provides a dedicated virtual channel to each paravirtualized OAI instance so that the OAI instances in different user slices are isolated from one another. 
    The virtual channels are created using the \textit{libvchan} interface, a Xen interface for inter-domain communication (IDC) that mainly relies on \texttt{gntalloc}, a userspace grant allocation driver that allocates shared pages, and \texttt{gntdev} that maps the granted pages into the address space of a userland process. 
   
  \item \textbf{Virtualization transparency:} 
  the OAI source code in user slices (i.e., DomUs) does not need to be changed in order to use PV-RAN. This is accomplished through \emph{API Remoting}, a software-based paravirtualization technique, that virtualizes the SDR at the programming API level. 
  In particular, the virtualization layer intercepts SDR function calls at run-time from the OAI instances running in DomUs, replaces them with API \textit{stubs}, and forwards them to Dom0 which in turn interacts with the physical SDR to actually execute the function calls.
\end{itemize}

In what follows, we elaborate on the PV-RAN virtualization layer. 
\ifthenelse{\boolean{9pager}}
{}
{ we also present the PV-RAN configurations in CyNet. }

\input{pvranImpl}

%% file: pvranImpl.tex

\subsection{PV virtualization layer}

\begin{figure}[!htbp]
    \centering
    \includegraphics[width=\linewidth]{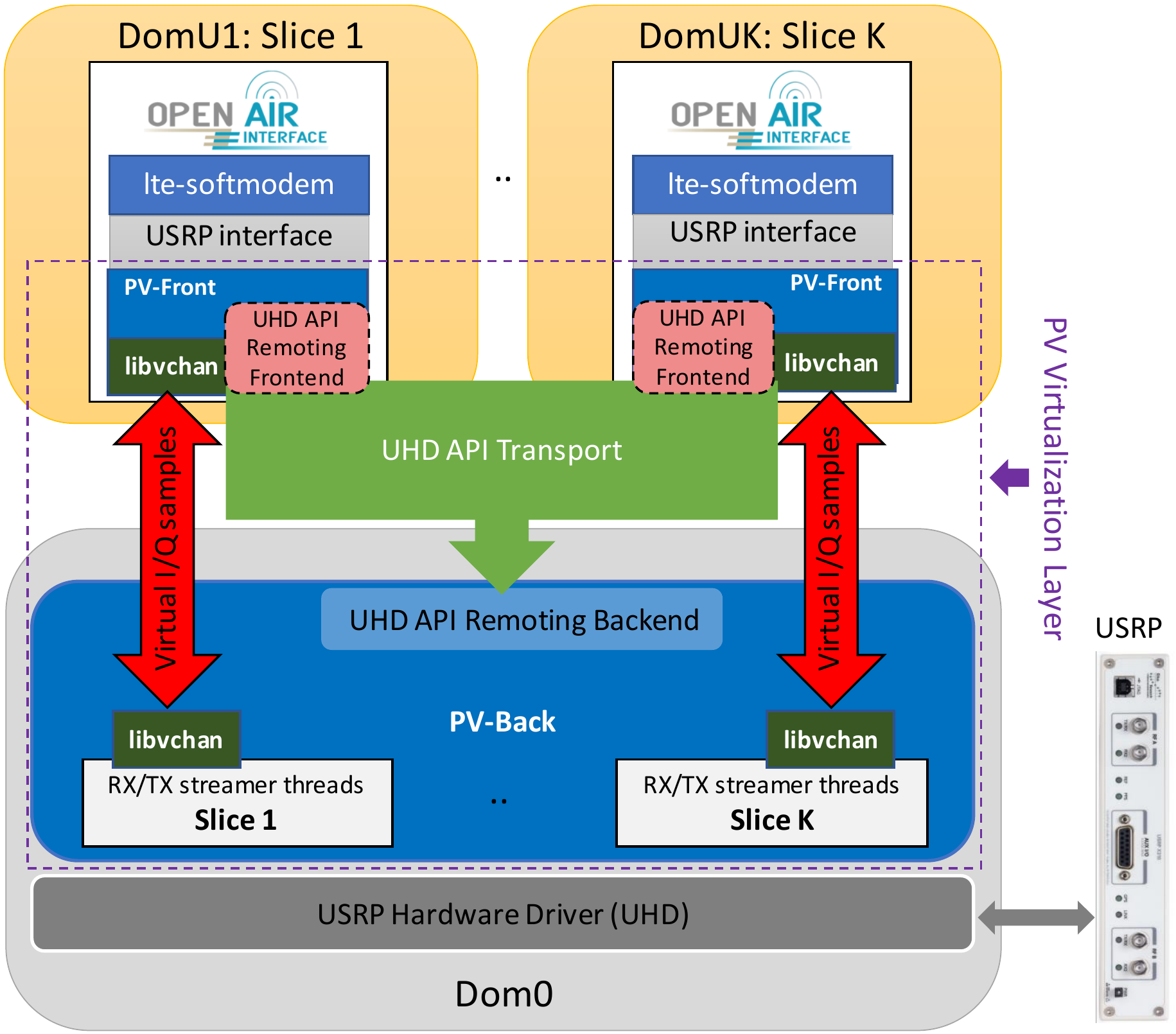}
    \caption{PV-RAN virtualization layer}
    \label{fig:PV-VL}
\end{figure}

Figure~\ref{fig:PV-VL} 
shows the internals of the PV virtualization layer. It consists of several cooperating modules located in DomU and Dom0. 
    In what follows, we first present API Remoting for transparent virtualization, and then we present PV-Back for SDR virtualization and inter-domain streaming of I/Q samples. 
    \ifthenelse{\boolean{9pager}}
    { (Information on more detailed implementation strategies can be found in the technical report \cite{pvran-TR}.) }
    {}

\subsubsection{Transparent virtualization via API Remoting}

OAI provides low-level interfaces to different types of SDR platforms such as HackRF, BladeRF, and USRP. In this paper we focus the discussion on a commonly used base station SDR --- USRP X310, but the design and implementation strategies can be applied to other SDRs. 
The USRP interface (usrp\_lib.cpp) is a software module that sits between the PHY layer and the SDR RF front-end. It implements functions that rely on UHD API functions to configure the USRP X310 (channel, bandwidth, gains, etc.) and perform RF I/O operations (i.e., send and receive time-domain I/Q samples). Since Dom0 interacts directly with the USRP X310 via UHD, all the UHD API calls in DomU  must be offloaded to Dom0. 
    One naive approach to addressing this challenge is to directly modify the source code of OAI in each DomU. While OAI is an open-source software allowing for modifications, a software-based mechanism that enables the  sharing of an SDR device among DomU guests without requiring OAI modification would be much more desirable. 
    To this end, we develop the API Remoting method that comprises the following modules:
\begin{itemize}
    \item \textit{UHD API Remoting Frontend:} this module is a dynamic library that relies on library interposition. It intercepts at run-time each UHD library function call and substitutes it with a \textit{stub} function or \textit{wrapper} that will execute another piece of code that forwards the call through the UHD API Transport to the UHD API Remoting Backend. The dynamic library is loaded during execution into the memory space associated with OAI lte-softmodem process. This is achieved through the use of the  \texttt{LD\_PRELOAD} environment variable that instructs the loader to load the dynamic library.
        Each stub makes a call to \texttt{dlsym()} with \texttt{RTLD\_NEXT} to locate the address of the related symbol in memory. Since OAI uses C++ bindings of the UHD API, we have to use mangled symbol names. Each UHD API C++ function has a mangled name that can be found in the dynamic symbol table. While the mangled symbol name is directly available for non-virtual C++ functions, the procedure to retrieve the mangled symbol name of virtual functions (e.g., UHD API setters and getters) is not trivial. It involves complex lookup in the virtual table which we will cover in our future work.
    
    \item \textit{UHD API Transport:} this transport component is a Xen event channel that allows the UHD API Remoting Frontend in PV-Front to send short control messages to the UHD API Remoting Backend in PV-Back. Each control message represents a UHD library function that must be executed by the UHD in Dom0. 
        For instance, upon start-up, the OAI instance in DomU (in particular, \textit{usrp\_lib.cpp}) will first issue the \texttt{uhd::device::find()} call to find the 
        USRP device connected. This call is then intercepted by the UHD API Remoting Frontend, and, upon interception, the stub calls another function that sends an \texttt{INIT} request to Dom0's UHD API Remoting Backend, which in turn 1) informs PV-Back in Dom0 to start streaming for this OAI slice and 2) returns to the UHD API Remoting Frontend in DomU the type of USRP device connected.
    
    \item \textit{UHD API Remoting Backend:} this module handles the call requests received over the UHD API Transport 
    and executes them on the SDR device using the actual UHD API.
        In its initial state, it waits for an incoming \texttt{INIT} request from a DomU's UHD API Remoting Frontend. This message is transmitted upon the execution of lte-softmodem in a DomU. The UHD API Remoting Backend processes the message and spawns RX/TX streamer threads which will be discussed shortly. 
\end{itemize}

\subsubsection{SDR virtualization through PV-Back in Dom0} 

Based on API remoting, PV-Back in Dom0 interacts with PV-Front in DomUs to perform RF I/O operations on behalf of paravirtualized OAI instances in different DomUs (i.e., user slices). 
It is designed to (i) initialize the USRP device,  configure the RX/TX frequencies, gains, bandwidths, and rates of the communication channel according to OAI base station (e.g., eNodeB) configuration in DomU, and map UHD RX/TX streamers to a dedicated SDR radio channel; 
    (ii) create bi-directional virtual channels and event channels for I/Q sample streaming and UHD API transport respectively; 
    and (iii) run high-priority RX/TX streamer threads for the continuous I/Q sample streaming between the USRP device and paravirtualized OAI instances in DomUs through the stream-based libvchan communication interface (to be discussed shortly). 

The RX/TX streamer threads are in charge of streaming I/Q samples. They are created on-demand upon the reception of an \texttt{INIT} request triggered when lte-softmodem is started in a DomU. Each paravirtualized OAI instance triggers the creation of two streamer objects: one RX streamer object that is the Dom0 interface to receive I/Q samples, and one TX streamer object that is the Dom0 interface to transmit I/Q samples. 
    PV-Back maps both streamers to one SDR radio channel per DomU and spawns two threads for the reception/transmission of I/Q samples.
Besides continuously receiving and transmitting I/Q samples, these two threads enable \emph{lightweight, in-band timestamping synchronization} between PV-Back in Dom0 and and PV-Front 
in DomU as follows: 
\begin{itemize}
    \item \textit{RX streamer thread:} it continuously reads \texttt{nsamples} (e.g., 7680 I/Q samples for 25-PRB channels) from the USRP device and then writes those samples to the virtual channel. During the first run, this thread generates a timestamp \texttt{rx\_timestamp} and transmits it 
    to PV-Front. 
    This timestamp is then used by OAI's  \texttt{trx\_usrp\_read} function as the time at which the first sample was received. (This timestamp is also used to compute the \texttt{tx\_timestamp} as we discuss shortly.) 
    After the first run, each time the thread performs an iteration of USRP device read, it uses the previous rx\_timestamp value and increases it by a fixed number depending on channel bandwidth (e.g., 7680 for 25-PRB  channels); so does that OAI instance (in particular, the \texttt{trx\_usrp\_read} function) in the corresponding DomU. 
        This way, the OAI instance in DomU and the TX streamer thread in Dom0 are synchronized in \texttt{rx\_timestamp} without any explicit coordination/messaging after the first run. 
    
    \item \textit{TX streamer thread:} it continuously reads \texttt{nbytes} (e.g., 30720 bytes for the 7680 I/Q samples for 25-PRB channels) from the virtual channel and then transmits those samples to the USRP device. During the very first run, this thread is locked and waits for  the RX streamer thread 
    to compute the \texttt{rx\_timestamp}. Once computed, the thread execution is resumed and it computes the appropriate \texttt{tx\_timestamp}, which is used by PV-Back to inform the UHD \texttt{send()} function at what time the first sample must be sent.
    The \texttt{tx\_timestamp} is computed considering the channel bandwidth. For instance, for 25-PRB channels, \texttt{tx\_timestamp} is computed as \texttt{rx\_timestamp + 30640}. 
    The OAI instance in the corresponding DomU (in particular, functions in lte-ru.c) also calculates, in the same way as the TX streamer thread in PV-Back, the \texttt{tx\_timestamp} based on the \texttt{rx\_timestamp} it receives from the RX streamer thread, and it uses it in the OAI code execution in DomU. This way, the TX streamer thread in Dom0 and the OAI instance in DomU are synchronized in \texttt{tx\_timestamp} without any explicit coordination/messaging. 
    Then, for each iteration of the virtual channel read, the thread  calculates the \texttt{tx\_timestamp} again by using the previous \texttt{tx\_timestamp} value and increasing it by a fixed number 
    (e.g., 7680 for 25-PRB channels).
\end{itemize}

\ifthenelse{\boolean{9pager}}
{}
{
Each of the aforementioned thread sets CPU affinity to run on a predefined core which makes it more convenient for performance monitoring.

The current implementation supports LTE bandwidths of 25 PRBs and 50 PRBs in FDD mode. We plan to incrementally add support for other features such as TDD mode and 100 PRBs bandwidth. The code of the PV-RAN implementation will be made available upon publication. 
} 

\subsubsection{Inter-domain streaming of I/Q samples} \label{subsec:idc}

PV-Back in Dom0 and PV-Front in DomUs create shared memory pages to communicate with each other. This functionality is available through the use of the grant table mechanism that allows kernel memory pages to be shared between 
domains. Each domain possesses a grant table which is a set of pages shared between itself and the Xen hypervisor. The domains involved in this mechanism are referred to as the client and the server, where the server offers the memory used for communication and transmits its credentials (grant reference, event channel ID) to the client in a Xen directory service known as XenStore. 
    The XenStore is a centralized database that stores key-value pairs and relies on the Linux kernel interface XenBus for messaging. To get access to the server's shared memory, the client needs the server's domain ID and the XenStore path in which the server offered its credentials. This phase is known as the rendezvous procedure.

Now, let's observe the mechanism in more detail by considering a DomU $A$, the client, wishing to communicate with another DomU $B$, the server:
\begin{itemize}
\item DomU $B$ creates a page entry in its grant table and then advertises the index of this entry (grant reference) to DomU $A$ through a dedicated Xen event channel. The kernel driver \textit{gntalloc} is the one in charge of creating grant references. 
\item Upon reception of the grant reference, DomU $A$ validates the grant and maps the page via the \textit{gntdev} device into the address space of the application running in DomU $A$.
\end{itemize}

Once the rendezvous procedure has been completed, Xen makes use of shared data structures called \textit{ring buffers} for bulk data transfer. Historically those ring buffers have been used by split drivers to communicate I/O requests-responses. 

In this paper we use Xen virtual channel library \textit{libvchan}, a library that implements a datagram-based (packet-based and stream-based) interface on top the standard Xen ring buffers. Libvchan allows to specify the size of the rings and whether or not to perform blocking  I/O operations on the virtual channel.
\ifshort 
\else 
    Listing~\ref{libxenvchan} shows details of the libvchan data structure. 
    \begin{lstlisting}[language=C,frame=single,caption=libvchan data structure,label=libxenvchan]
    struct libxenvchan_ring {
      struct ring_shared* shr;
      void* buffer;
      int order;
    }
    
    struct libxenvchan {
      union {
        xengntshr_handle *gntshr;
        xengnttab_handle *gnttab;
      };
      struct vchan_interface *ring;
      xenevtchn_handle *event;
      uint32_t event_port;
      int is_server:1;
      int server_persist:1;
      int blocking:1;
      struct libxenvchan_ring read, write;
    };
    \end{lstlisting} 
    Each \textit{libxenvchan} data structure contains information about the role of the structure (client or server), the event channel interface and its port number, a pointer to a shared ring page \textit{vchan\_interface}, two \textit{libxenvchan\_ring} communication ring variables  \textit{read} and \textit{write} that are composed of a pointer into the shared page and a ring data buffer, the mapping handle for shared ring page (\textit{gntshr} for server and \textit{gnttab} for client), and whether or not the virtual channel performs blocking I/O operations.

\fi
As OAI deals with streams of I/Q samples, we use the stream-based communication interface that is composed of  \textit{libxenvchan\_read} and \textit{libxenvchan\_write} functions to read I/Q samples from a buffer and write I/Q  samples to a buffer respectively. The blocking I/O flag is set on the bi-directional virtual channel to ensure that I/O operations are performed sequentially.


\ifthenelse{\boolean{9pager}}
{}
{ \input{pvranInCyNet} }

%% file: pvranInCyNet.tex
\subsection{System configuration details in CyNet} 

Table~\ref{tab:hwsoft} 
\begin{table}[!htbp]
\centering
\begin{tabular}{ |p{2cm}||p{4cm}| }
 \hline
 \multicolumn{2}{|c|}{\textbf{Hardware}} \\
 \hline
 System & Dell Tower Precision 3630 \\
 \hline
 CPU &  Intel Core i9-9900 @ 3.10GHz\\
 \hline
 Memory & 64 GB\\
 \hline
 Network & Intel Ethernet Adapter x722-DA4 \\
 \hline
 SFP+ Transceiver & Intel FTLX1471D3BCV-I3 SFP+ 10G 1310nm \\
 \hline
 SDR & USRP x310 + two UBX-160 daughterboards \\
 \hline
 \hline
 \multicolumn{2}{|c|}{\textbf{Xen Hypervisor}} \\
 \hline
 Hypervisor & Xen 4.9.2 \\
 \hline
 Dom0 OS & Ubuntu 18.04.4 LTS \\
 \hline
 Volume group size & 475 GB \\
 \hline
 UHD driver & v3.15.0.0 \\
 \hline
 Intel NIC driver & i40e version 2.1.14-k \\
 \hline
 \hline
 \multicolumn{2}{|c|}{\textbf{Xen DomU Guests}} \\
 \hline
 DomU OS & Ubuntu 16.04.4 LTS \\
 \hline
 vCPUs & 4 \\
 \hline
 Memory & 6 GB\\
 \hline
 Logical volume size & $>$ 18GB \\
 \hline
 Xen shared libraries & libxenvchan, libxenevtchn, libxengnttab, libxentoolcore, libxentoollog\\
 \hline
 OpenAirInterface & v1.2.1 \\
 \hline
 UHD driver & v3.15.0.0 \\
 \hline
\end{tabular}
\caption{Hardware and software configurations in CyNet}
\label{tab:hwsoft}
\end{table}
summarizes the hardware and software configurations in CyNet. In what follows, we elaborate on the configurations of Dom0, DomU, and the fronthaul link between the SDR and BS computer. 

\subsubsection{Dom0 configuration} 

The PV-RAN runs under a Dell Precision 3630 equipped with an Intel Core i9-9900 with 8 physical cores running at 3.1 GHz, and 64 GB of RAM. Since OAI has been developed and optimized for Ubuntu, the Dell desktop uses Ubuntu 18.04.4 (bionic) with Xen hypervisor version 4.9.2. 
Xen Dom0 uses its own physical HDD while DomUs sit in another HDD in LVM-managed logical volumes that can be easily resized if need be.

\subHeading{Dom0 low-latency \& IDC kernel configuration.}
As OAI must deal with real-time deadlines, a custom 4.15.18 Linux kernel has been installed on Dom0 with kernel features necessary to support  time-critical applications. Thus, we have enabled \texttt{PREEMPT} support, set \texttt{HZ} (timer resolution) to 250 Hz, and enabled \texttt{CONFIG\_HIGH\_RES\_TIMERS} to achieve low latency timers. The reason we do not use a prebuilt low-latency kernel is that IDC requires to enable Xen-specific kernel modules such as gntalloc, gntdev, evtchn and xenbus.

\subHeading{Dom0 CPU frequency scaling.}
OAI is a demanding software with real-time requirements. Therefore, it is essential to tune the system adequately. Power-savings mechanisms can force the processor to transition to low-power states where the CPU stops executing instructions to reduce the power consumption. When the CPU is woken up by an instruction, it will switch to a lower power saving state. When the CPU transitions from one power state to another it results in some extra wake-up latency which degrades the performance of real-time applications. Since Dom0 performs all the periodic RF I/O operations between the SDR platform and the paravirtualized instances of OAI in DomUs, the wakeup latency must be minimized. Thus, it is highly recommended to disable all power management features (C-states, P-states).

On one hand, disabling C-states (or core power states) means that the CPU will never be in idle sleep state where its clock is inactive. Hence, the CPU will always run in C0 state which will result in low-latency wakeup. On the other hand, disabling P-states means that the CPU will always run at the highest frequency (and highest voltage) in P0 state. To force the CPU to remain in C0 and P0 state, we disable dynamic frequency scaling also known as \textit{ondemand governor}, usually loaded by default in order to save power when the CPU is idle. Those cpufreq policies are managed by a CPU frequency driver \textit{cpufreq governor} in the Linux kernel. The procedure to tune CPU frequency scaling for OAI in a bare-metal system is well-documented. However, in a Xen environment, we have to enable Dom0-based cpufreq which lets  Dom0 handle the cpufreq logic. This is done by adding the option \texttt{cpufreq=dom0-kernel} in Xen's grub bootloader entry. Finally, the Dom0's userspace control tool \texttt{xenpm} is used to select the performance cpufreq governor, to force the processor to stay in C0 and to enable the turbo mode on all CPUs.

\subsubsection{DomU configuration}

Each DomU guest relies on Ubuntu 16.04 LTS with a 4.9.214 Linux kernel compiled with Xen modules necessary for IDC (gntalloc, gntdev, evtchn, xenbus). Since this is where the paravirtualized LTE protocol stack runs, we installed OAI v1.2.1 from OAI's gitlab master branch. OAI's build system requires the path to the current kernel's build directory. Since DomUs use a custom kernel, the build directory is not located in /lib/modules but in /usr/src, so a symbolic link must be set before building OAI. Dom0 allocates each DomU with 4 vCPUs and 6 GB of memory. Each DomU's LVM logical disk requires a minimum size of 18 GB for the OS, OAI, UHD and their dependencies.

\subsubsection{High-speed optical fronthaul link}

The PV-RAN uses a high-performance SDR platform from NI/Ettus, the USRP X310, equipped with two daughter boards UBX-160 and a high-speed interface of 10Gbps which is connected to the Dell desktop described earlier and fully managed by Dom0's native network driver. An OS2 single-mode optical fiber connects the Dell desktop and the USRP X310. Each side of the fiber is plugged to an SFP+ Intel transceiver module. This link is in charge of transferring I/Q samples between the host (Dell desktop) and the SDR platform. 
    The LTE branch of OAI supports bandwidths of 5, 10 and 20MHz. OAI uses 16 bits to represent each sample, that is 16 bits for I (In-Phase) and 16 bits for Q (Quadrature). Therefore, a 1Gbps Cat-6 copper cable would be enough to accommodate LTE bandwidths up to 20 MHz as described in Table~\ref{tab:linkspeed}. 
\begin{table}[!htbp]
\centering
\caption {Link speed for different OAI LTE bandwidths} \label{tab:linkspeed} 
\resizebox{.45\textwidth}{!}{
\begin{tabular}{|l|l|l|l|}
    \hline
    \textbf{Bandwidth} & \textbf{PRBs} & \textbf{Sample rate} & \textbf{Required link speed}\\
    \hline
    5 MHz & 25 & 7.68 Msps & 245.76 Mbit/s\\
    10 MHz & 50 & 15.36 Msps & 491.52 Mbit/s\\
    20 MHz & 100 & 30.72 Msps & 983.94 Mbit/s\\
    \hline
\end{tabular}}
\end{table}
However, since the PV-RAN has been deployed in the context of the CyNet project where the RRH is located on Ames' Research Park rooftop more than 100 feet from the Dell desktop (located in Research Park's basement), 10Gbps optical fiber is used between the host and the USRP X310. Using optical fiber as the fronthaul link has many benefits over copper cable: higher sample rates (up to 200Msps) can be enabled for use cases with wider channel bandwidths, stronger resistance to electromagnetic interference and radio signals, and a weaker signal degradation over long distances.

%% file: measurement.tex
\section{Measurement Evaluation} \label{sec:evaluation}

We have implemented the PA-RAN design \cite{pvran-TR} and integrated it with the CyNet wireless living lab \cite{CyNet}. A recorded demo of the operational PV-RAN can be found at \cite{pvran-demo}. 
    Here we conduct detailed measurement benchmarking to characterize the performance of the PV-RAN platform.

\subsection{PV-RAN 
testbed}

\begin{figure}[!htbp]
    \centering
    \includegraphics[width=.8\linewidth]{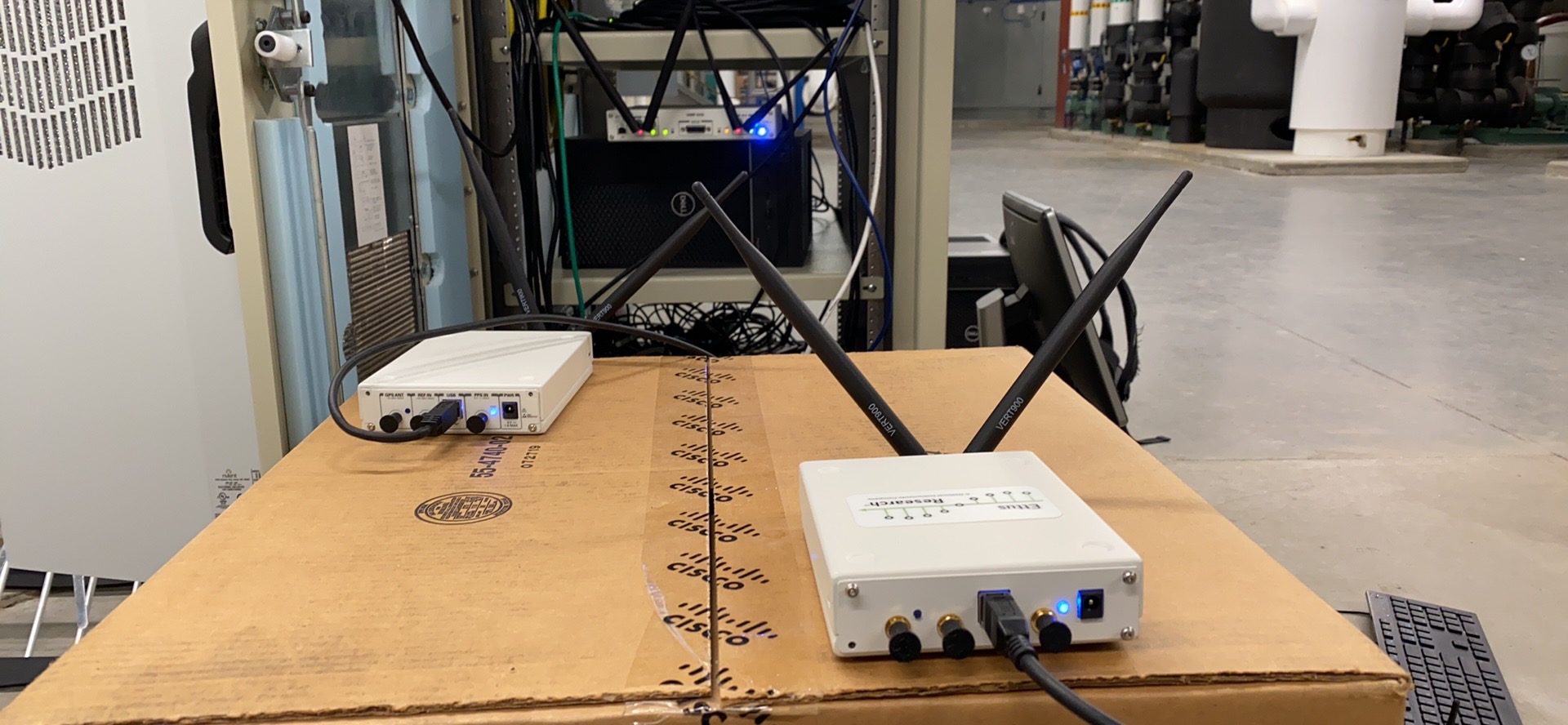}
    \caption{PV-RAN testbed at ISU Research Park} 
    \label{fig:testbed}
    \vspace*{-0.1in}
\end{figure}

We set up a 
testbed in the basement of the Iowa State University (ISU) Research Park to conduct our measurement evaluation (see Figure~\ref{fig:testbed}), with the basement having significantly less interference from surrounding wireless equipment. The testbed is composed of:
\begin{itemize}
    \item One PV-RAN platform running two slices: the base station computer hosting PV-RAN under the Xen hypervisor is a Dell Precision 3630 with 64GB of RAM and 8 CPU cores operating at 3.1GHz. Each DomU slice has been set up as a paravirtualized guest under Ubuntu 16.04 and is configured with 4 vCPUs and 6GB of RAM. Each DomU slice runs OpenAirInterface as the LTE protocol stack. Bi-directional virtual channels transport I/Q samples between each DomU PV-Front and Dom0 PV-Back. Dom0 runs under Ubuntu 18.04 and has direct I/O access to the USRP X310 which is connected to the base station computer through OS2 single-mode fiber. 
    \item Two UEs:  each UE runs on an Intel NUC7i7BNB under Ubuntu 16.04 with a low-latency kernel. Each NUC is connected via USB 3.0 to a USRP B210 equipped with two VERT900 antennas that can operate in both the Educational Broadband Service (EBS) and TV White Space (TVWS) frequencies.
\end{itemize}

\subsection{Network performance tuning}

The USRP X310 is connected via single-mode fiber to one of the SFP+ network interface of the base station computer which provides a 10Gbps network speed. 
Table~\ref{tab:netif} summarizes the parameters that must be fine-tuned to allow the best network performance possible. We used ifconfig and ethtool to configure the network interface. It is recommended to use jumbo frames (for larger-packet transmission support), to increase the size of RX/TX kernel buffers and Ethernet buffer rings, to disable Ethernet frame pauses (RX/TX autoneg off), and to minimize the raise of interrupts to the CPU during packet reception to improve CPU utilization.

\begin{table}[!htbp]
\centering
\begin{tabular}{ |p{3cm}||p{2cm}| }
 \hline
 \multicolumn{2}{|c|}{\textbf{Network interface parameters}} \\
 \hline
 MTU size & 8000 bytes  \\
 \hline
 Queue length &  1000 bytes \\
 \hline
 RX/TX autoneg & off \\
 \hline
 Ethernet buffer rings & 4096 bytes \\
 \hline
 Interrupt moderation & 3 $\mu$secs \\
 \hline
\end{tabular}
\caption{Hardware and software setup}
\label{tab:netif}
\vspace*{-0.1in}
\end{table}


\subsection{Throughput and Latency} 

To assess the smooth operation of PV-RAN, we compare the throughput of one PV-RAN slice running OAI against a non-virtualized instance of OAI running in a bare-metal server. The bare-metal server running OAI has the same specification as the base station computer used for  PV-RAN. Both the PV-RAN OAI and Bare-metal OAI use the same LTE bandwidth of 25 PRBs (5MHz) and operate at the LTE downlink frequency and uplink frequency of 2.685GHz and 2,565GHz respectively. The main eNB parameters  are shown in Table~\ref{tab:thr-eval} along with UE parameters. 
\begin{table}[!htbp]
\centering
    \begin{tabular}{|c||c|}
    \hline
    \multicolumn{2}{|c|}{\textbf{eNB configuration}} \\
    \hline
    LTE bandwidth & 25 PRBs\\
    \hline
    Downlink frequency & 2.685 GHz\\
    \hline
    Uplink frequency & 2.565 GHz\\
    \hline
    PDSCH reference signal power & -16 dB \\
    \hline
    \hline
    \multicolumn{2}{|c|}{\textbf{UE configuration}} \\
    \hline
    RX gain & 120 dB \\
    \hline
    TX gain & 0 dB \\
    \hline
    Max UE power & 0 dB \\
    \hline
    \end{tabular}
    \caption{eNB and UE configuration}
    \label{tab:thr-eval}
    \vspace*{-0.1in}
\end{table}

\begin{figure}[!htbp]
    \centering
    \includegraphics[width=.8\linewidth]{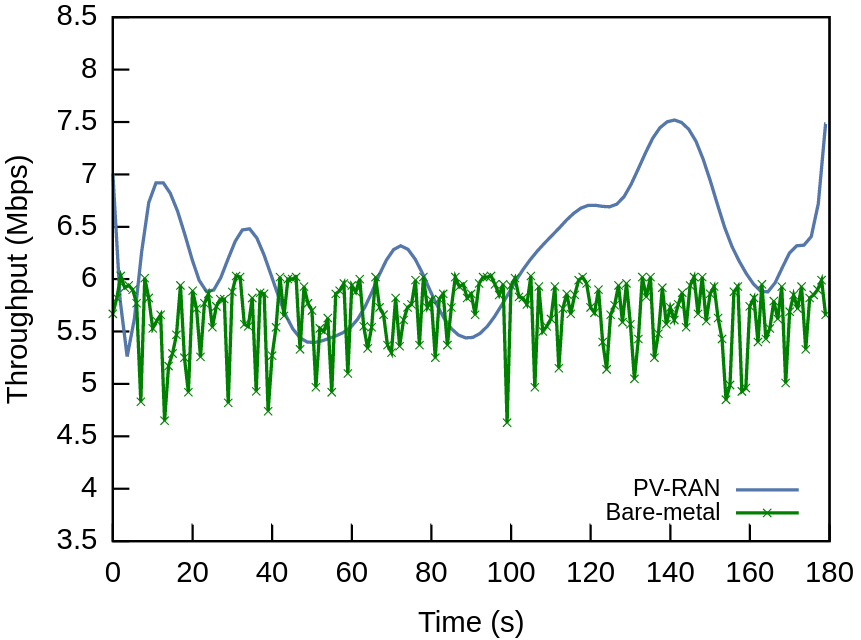}
    \caption{Throughput comparison between PV-RAN OAI and Bare-metal OAI}
    \label{fig:thr-virt-nonvirt}
    \vspace*{-0.1in}
\end{figure}

We attach one UE to each eNB and measure the UDP throughput for 180 seconds using the iperf tool. 
The throughput in the PV-RAN OAI and bare-metal OAI are 
similar. For instance, 
Figure~\ref{fig:thr-virt-nonvirt}
depicts the throughput between each eNB and their associated UE in one single run. The mean throughput in the PV-RAN OAI and bare-metal OAI are 6.24Mbps and 5.66Mbps, respectively; the difference is mostly due to environmental and wireless channel dynamics. 

\begin{figure}[!htbp]
    \centering
    \includegraphics[width=.8\linewidth]{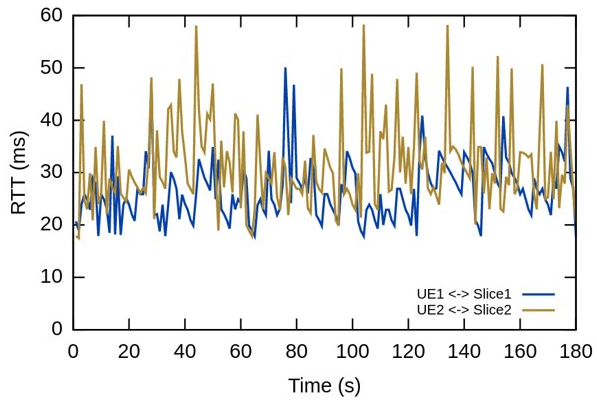}
    \caption{RTT for each PV-RAN OAI slice}
    \label{fig:rtt-dual}
    \vspace*{-0.1in}
\end{figure}

\begin{figure}[!htbp]
    \centering
    \includegraphics[width=.8\linewidth]{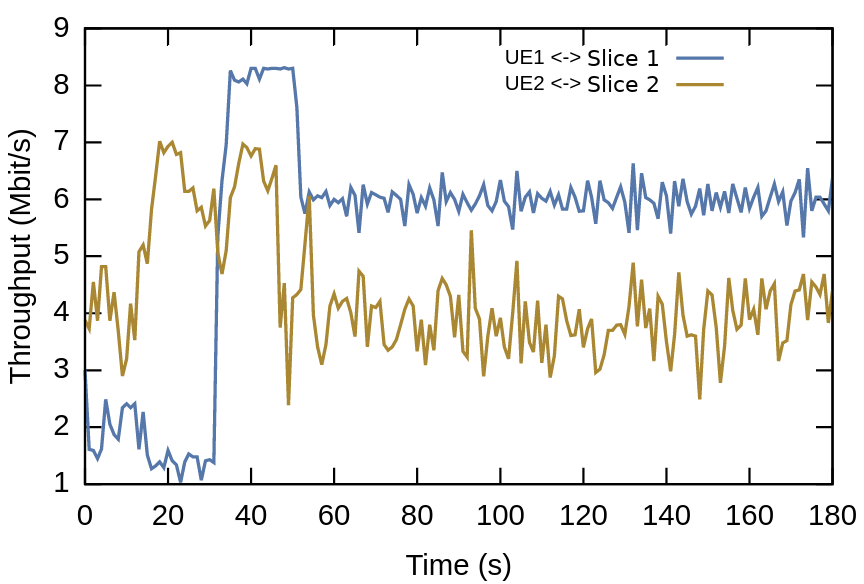}
    \caption{UDP Throughput for each PV-RAN OAI slice}
    \label{fig:thr-dual}
    \vspace*{-0.1in}
\end{figure}

Next, we evaluate the performance when two OAI slices are running on PV-RAN. In this experiment, we measure the RTT and throughput between the OAI eNB and UE in each slice. Each OAI eNB uses a channel bandwidth of 25 PRBs. 
    Both slices operate in TVWS bands: slice 1 runs at 595MHz (downlink) and 499MHz (uplink), and slice 2 at 580MHz (downlink) and 484MHz (uplink). Each slice occupies a different radio channel on the USRP X310. We connect one UE to each slice and measure the RTT with the ping command and the UDP throughput with the iperf command. The ICMP echo requests have been sent from each UE to their respective eNB. 
Figure~\ref{fig:rtt-dual} 
illustrates the RTT variation over time for each UE-eNB pair. The average RTT is of 25ms for slice 1 and 29ms for slice 2. We witness similar trends with OAI running on bare-metal servers. 
    The throughput of each OAI slice is depicted in Figure~\ref{fig:thr-dual}.
It can be seen that slice 1 achieves a stable throughput of $\sim$6Mbps on average while slice 2 achieves $\sim$4Mbps on average, which are also comparable to throughput achieved with bare-metal OAI. The reason why slice 1 has slightly lower RTT and higher throughput than slice 2 is because the UE in slice 1 is closer to its eNB and thus have better channel quality. (The slice throughput is less stable at the beginning phase when UEs and eNBs dynamically tune their communication parameters such as transmission power and MCS schemes.)


\subsection{CPU utilization}

It is of paramount importance to understand the CPU consumption of our PV-RAN platform. For this study, all CPU cores ran at their maximum frequency in Dom0. Each slice spawns two threads in PV-Back, namely \textit{stream-rx} that receives I/Q samples from the USRP device and writes them to the virtual channel, and \textit{stream-tx} that reads I/Q samples from the virtual channel and sends them to the USRP device. We pinned each thread onto a single CPU core using \texttt{pthread\_setaffinity\_np} to have a better understanding of CPU usage at a thread level. 
    In this experiment, the useful Linux task monitoring tools \texttt{top} and \texttt{htop} have been used to collect CPU consumption of individual threads. 

\begin{figure}[!htbp]
    \centering
    \includegraphics[width=.8\linewidth]{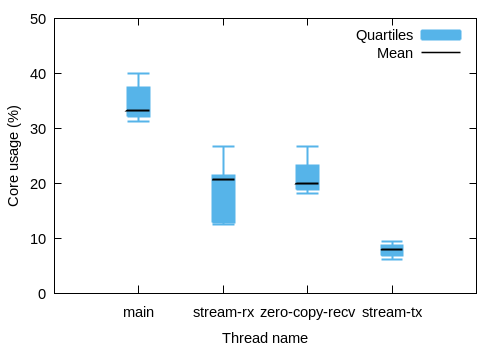}
    \caption{Core usage per thread for Dom0 PV-Back}
    \label{fig:core-usage}
    \vspace*{-0.1in}
\end{figure}

The CPU core usage per thread is reported in Figure~\ref{fig:core-usage} 
for a single OAI slice running on PV-RAN. The \textit{main} thread listens for incoming PV-Front requests and is responsible for the creation of USRP device object and virtual channel data structures. It has a mean CPU core consumption of 33.3\%. It can be observed that an additional thread named \textit{zero-copy-recv} is spawned by the call to the UHD \textit{recv()} function in the \textit{stream\_rx} thread. This thread is one of the most influential thread in terms of CPU consumption for OAI eNBs. It helps to minimize unnecessary buffer copy operations between the NIC and userspace by bypassing buffer copy operations from the NIC to the kernel space and from the  kernel space to the userspace. \textit{stream-rx} and \textit{zero-copy-recv} consume 20.7\% and 20\% respectively. These tasks play a major role in CPU consumption, while in contrast the thread \textit{stream-tx} has a mean consumption of only 8\% of its CPU core.

\begin{figure}[!htbp]
    \centering
    \includegraphics[width=.8\linewidth]{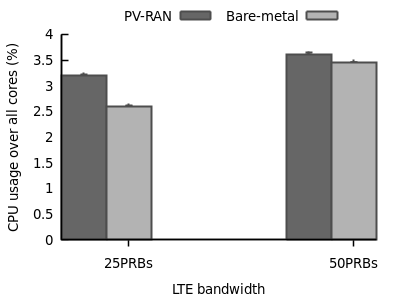}
    \caption{Mean CPU usage over all cores for different bandwidths}
    \label{fig:meancpu}
    \vspace*{-0.1in}
\end{figure}

We also performed a comparative analysis of the total CPU consumption (over all CPU cores) between OAI running in bare-metal and OAI running as a PV-RAN slice. The CPU usage over all cores is depicted in Figure~\ref{fig:meancpu}. 
It can be seen that, for a LTE bandwidth of 25 PRBs, OAI eNB in bare-metal consumes about 2.6\% of CPU, and it consumes 3.2\% of CPU when running as a PV-RAN slice, a slight increase over bare-metal operation. 
    When the LTE channel bandwidth is doubled (i.e., with 50 PRBs), the CPU usage for OAI in bare-metal increases by 0.85\% while the CPU usage for OAI in PV-RAN increases by only 0.4\%. So the CPU usage difference between bare-metal and PV-RAN operations actually decreases with increasing channel bandwidth. 
Therefore, the CPU usage overhead due to virtualization in PV-RAN is small, and tends to be negligible as channel bandwidth increases. 




\subsection{Latency overhead }

\begin{figure}[!htbp]
    \centering
    \includegraphics[width=.8\linewidth]{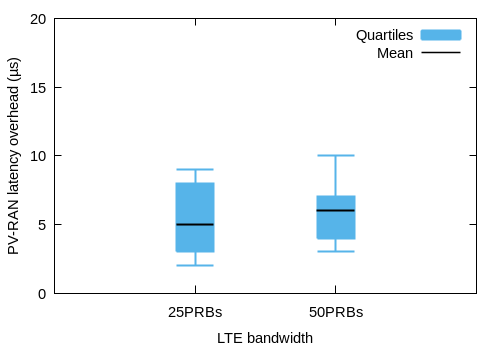}
    \caption{Latency overhead of inter-domain I/Q sample streaming}
    \label{fig:overhead-latency}
    \vspace*{-0.1in}
\end{figure}

To complete our measurement study, we focus on latency overhead which, if not controlled to be low, can be detrimental to I/Q sample streaming and cellular communication performance. Since we add an extra layer between OAI USRP interface and UHD I/O functions to transfer I/Q samples, we need to evaluate the latency overhead introduced by the PV virtualization layer. For this case study, in particular, we add timestamps in PV-Front and PV-Back to measure the latency overhead caused by the virtual channels.
Figure~\ref{fig:overhead-latency}
shows the latency overhead in PV-RAN. 
 Surprisingly, for a channel bandwidth of 25 PRBs, the mean overhead introduces by the virtual channel is small and only 5$\mu$s on average. For a channel bandwidth of 50 PRBs which doubles the ring buffer size in the virtual channel and tends to introduce higher latency, the latency overhead is also only 6$\mu$s on average. 
This low latency overhead and high efficiency of PV-RAN explain the good throughput performance of PV-RAN discussed earlier. 

Based on the above latency results, we decide to evaluate the theoretical upper-bound capacity of our PV-RAN platform. More specifically, we verify how the inter-domain I/Q sample streaming scales with respect to the number of OAI slices for different LTE bandwidths. 
    In addition, the Wireless Spectrum Hypervisor~\cite{de2020baseband} uses a ZMQ socket with the PUB-SUB (Publish-Subscribe) pattern for I/Q sample streaming. 
    Therefore, we also compare the PV-RAN method of shared-memory inter-domain I/Q sample streaming against the ZMQ PUB-SUB I/Q sample streaming method. To this end, we create a variant of PV-RAN that uses a ZMQ socket to  publish I/Q samples between PV-Front and PV-Back. 
    
\begin{figure}[!htbp]
    \vspace*{-0.05in}
    \centering
    \includegraphics[width=.76\linewidth]{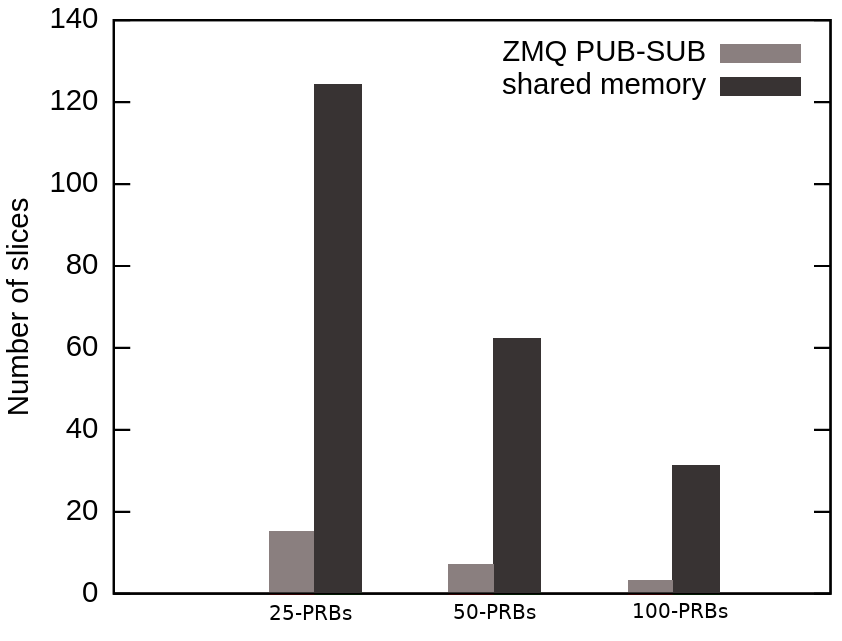}
    \caption{OAI slice scalability for different LTE bandwidths}
    \label{fig:scalability}
    \vspace*{-0.1in}
\end{figure}

The latency overhead observed for ZMQ PUB-SUB varies between 65$\mu$s and 90$\mu$s, significantly higher than that with the PV-RAN method. From these results we can infer the maximum capacity for ZMQ PUB-SUB between Dom0 and DomU too. Figure~\ref{fig:scalability} 
shows for different LTE bandwidths that shared-memory clearly outperforms ZMQ PUB-SUB, with shared-memory inter-domain I/Q sample streaming being 7 times faster than ZMQ PUB-SUB inter-domain I/Q sample streaming on average. While ZMQ PUB-SUB remains a viable solution for general communications between processes on the same host, it is not a good candidate for intensive real-time inter-domain I/Q sample streaming.





%% file: conclusion.tex
\section{Concluding Remarks} \label{sec:conclusion}

As the first open-source platform for physical wireless resource virtualization, PV-RAN enables whole-stack slicing where different PHY and MAC layers may be adopted for diverse communication services and wireless living lab innovations. 
    Using efficient techniques such as API Remoting, shared-memory I/Q sample streaming, and light-weight in-band timestamping synchronization, PV-RAN enables transparent, lightweight virtualization of SDRs, and it serves a solid foundation for further development. 
For instance, the PV-RAN platform can be extended to support SDR virtualization using TDM and hybrid-TDM-FDM, and the PV-RAN implementation can be ported to other open-source platforms such as KVM and srsLTE to expand community access to PV-RAN capabilities. 

%% file: ms.bbl
\begin{thebibliography}{10}
\providecommand{\url}[1]{#1}
\csname url@samestyle\endcsname
\providecommand{\newblock}{\relax}
\providecommand{\bibinfo}[2]{#2}
\providecommand{\BIBentrySTDinterwordspacing}{\spaceskip=0pt\relax}
\providecommand{\BIBentryALTinterwordstretchfactor}{4}
\providecommand{\BIBentryALTinterwordspacing}{\spaceskip=\fontdimen2\font plus
\BIBentryALTinterwordstretchfactor\fontdimen3\font minus
  \fontdimen4\font\relax}
\providecommand{\BIBforeignlanguage}[2]{{%
\expandafter\ifx\csname l@#1\endcsname\relax
\typeout{** WARNING: IEEEtran.bst: No hyphenation pattern has been}%
\typeout{** loaded for the language `#1'. Using the pattern for}%
\typeout{** the default language instead.}%
\else
\language=\csname l@#1\endcsname
\fi
#2}}
\providecommand{\BIBdecl}{\relax}
\BIBdecl

\bibitem{alvarez2019edge}
F.~Alvarez, D.~Breitgand \emph{et~al.}, ``{An edge-to-cloud virtualized
  multimedia service platform for 5G networks},'' \emph{IEEE Transactions on
  Broadcasting}, vol.~65, no.~2, pp. 369--380, 2019.

\bibitem{zhang2017network}
H.~Zhang, N.~Liu, X.~Chu, K.~Long, A.-H. Aghvami, and V.~C. Leung, ``Network
  slicing based 5g and future mobile networks: mobility, resource management,
  and challenges,'' \emph{IEEE Communications Magazine}, vol.~55, no.~8, pp.
  138--145, 2017.

\bibitem{peng2019packet}
S.~Peng, R.~Chen, and G.~Mirsky, ``Packet network slicing using segment
  routing,'' Tech. Rep. Draft-penglsr-network-slicing-00. IETF, Tech. Rep.,
  2019.

\bibitem{FlexRAN}
``{}flexran: A flexible and programmable platform for software-defined radio
  access networks.''

\bibitem{esmaeily2020cloud}
A.~Esmaeily, K.~Kralevska, and D.~Gligoroski, ``{A Cloud-based SDN/NFV Testbed
  for End-to-End Network Slicing in 4G/5G},'' \emph{arXiv:2004.10455}, 2020.

\bibitem{wirelessVirtualization:survey}
M.~Yang, Y.~Li, D.~Jin, L.~Zeng, X.~Wu, and A.~V. Vasilakos,
  ``{Software-Defined and Virtualized Future Mobile and Wireless Networks: A
  Survey},'' \emph{Mobile Networks and Applications}, vol.~20, 2015.

\bibitem{CyNet}
``{CyNet: End-to-End Software-Defined Cyberinfrastruture for Smart Agriculture
  and Transportation},''
  \url{https://www.ece.iastate.edu/~hongwei/group/projects/CyNet.html}.

\bibitem{UCS-ICII18}
Y.~Xie, H.~Zhang, and P.~Ren, ``Unified scheduling for predictable
  communication reliability in industrial cellular networks,'' in \emph{IEEE
  ICII}, 2018.

\bibitem{USRP}
E.~Research, ``{USRP Software-Defined Ratios},''
  \url{https://www.ettus.com/product}.

\bibitem{Xen}
P.~Barham, B.~Dragovic, K.~Fraser, S.~Hand, T.~Harris, A.~Ho, R.~Neugebauer,
  I.~Pratt, and A.~Warfield, ``{Xen and the Art of Virtualization},'' in
  \emph{ACM SOSP}, 2003.

\bibitem{OAI}
N.~Nikaein, M.~K. Marina, S.~Manickam, A.~Dawson, R.~Knopp, and C.~Bonnet,
  ``{OpenAirInterface: A Flexible Platform for 5G Research},'' \emph{ACM
  SIGCOMM Computer Communication Review}, vol.~44, no.~5, pp. 33--38, 2014.

\bibitem{suryaprakash2015heterogeneous}
V.~Suryaprakash, P.~Rost, and G.~Fettweis, ``Are heterogeneous cloud-based
  radio access networks cost effective?'' \emph{IEEE Journal on Selected Areas
  in Communications}, vol.~33, no.~10, pp. 2239--2251, 2015.

\bibitem{garcia2018fluidran}
A.~Garcia-Saavedra, X.~Costa-Perez, D.~J. Leith, and G.~Iosifidis, ``{FluidRAN:
  Optimized vran/mec orchestration},'' in \emph{IEEE INFOCOM}, 2018.

\bibitem{nikaein2017towards}
N.~Nikaein, E.~Schiller, R.~Favraud, R.~Knopp, I.~Alyafawi, and T.~Braun,
  ``Towards a cloud-native radio access network,'' in \emph{Advances in mobile
  cloud computing and big data in the 5G era}.\hskip 1em plus 0.5em minus
  0.4em\relax Springer, 2017, pp. 171--202.

\bibitem{kaltenberger2020openairinterface}
F.~Kaltenberger, A.~P. Silva, A.~Gosain, L.~Wang, and T.-T. Nguyen,
  ``Openairinterface: Democratizing innovation in the 5g era,'' \emph{Computer
  Networks}, p. 107284, 2020.

\bibitem{tran2017understanding}
T.~X. Tran, A.~Younis, and D.~Pompili, ``Understanding the computational
  requirements of virtualized baseband units using a programmable cloud radio
  access network testbed,'' in \emph{IEEE ICAC}, 2017.

\bibitem{sanchez20205gtoponet}
I.~Sanchez-Navarro, A.~S. Mamolar, Q.~Wang, and J.~M.~A. Calero, ``5gtoponet:
  Real-time topology discovery and management on 5g multi-tenant networks,''
  \emph{Future Generation Computer Systems}, vol. 114, pp. 435--447, 2020.

\bibitem{trindade2019c}
I.~Trindade, C.~Nahum, C.~Novaes, D.~Cederholm, G.~Patra, and A.~Klautau,
  ``C-ran virtualization with openairinterface,'' \emph{arXiv preprint
  arXiv:1908.07503}, 2019.

\bibitem{nikaein2015processing}
N.~Nikaein, ``Processing radio access network functions in the cloud: Critical
  issues and modeling,'' in \emph{ACM MCS}, 2015.

\bibitem{de2020baseband}
F.~A. de~Figueiredo, R.~Mennes, I.~Jaband{\v{z}}i{\'c}, X.~Jiao, and
  I.~Moerman, ``A baseband wireless spectrum hypervisor for multiplexing
  concurrent ofdm signals,'' \emph{Sensors}, vol.~20, no.~4, p. 1101, 2020.

\bibitem{3gpp-functionalSplit}
``Study on new radio access technology: Radio access architecture and
  interfaces,'' \emph{3GPP TR38.801 Release 14}.

\bibitem{LightVM}
F.~Manco, J.~Mendes, K.~Yasukata, C.~Lupu, S.~Kuenzer, C.~Raiciu, F.~Schmidt,
  S.~Sati, and F.~Huici, ``{My VM is Lighter (and Safer) than your
  Container},'' in \emph{ACM SOSP}, 2017.

\bibitem{pvran-TR}
M.~Sander-Frigau, T.~Zhang, H.~Zhang, A.~E. Kamal, and A.~K. Somani, ``Physical
  wireless resource virtualization for software-defined whole-stack slicing,''
  Iowa State University, Tech. Rep. ISU-DNC-TR-20-02
  (\url{https://www.ece.iastate.edu/~hongwei/group/publications/PV-RAN-TR.pdf}),
  2020.

\bibitem{pvran-demo}
``{PV-RAN Demo},''
  \url{https://www.ece.iastate.edu/~hongwei/group/projects/CyNet/PV-RAN-CyNet-Demo.mp4}.

\end{thebibliography}
